\documentclass[twocolumn,aps,prd,epsf]{revtex4}
\usepackage{epsfig}

\begin{document}
\title{ Analytical approach to chiral symmetry breaking in Minkowsky space }
\author{Pedro Bicudo}
\address{Departamento de F\'{\i}sica and \ 
Centro de F\'{\i}sica das Interac\c c\~oes Fundamentais, \\
Edif\'{\i}cio Ci\^encia, 
Instituto Superior T\'ecnico, Av. Rovisco Pais, 
1049-001 Lisboa, Portugal}
\begin{abstract}
The mass gap equation for spontaneous chiral symmetry breaking is studied 
directly in Minkowsky space. 
In hadronic physics, spontaneous chiral symmetry breaking is crucial to 
generate a constituent mass for the quarks, and to produce 
the Partially Conserved Axial Current theorems, including a small mass 
for the pion.
Here a class of finite kernels is used, expanded in Yukawa interactions. 
The Schwinger-Dyson equation is solved with an analytical approach. 
This improves the state of the art of solving the mass gap 
equation, which is usually solved with the equal-time approximation or 
with the Euclidean approximation.
The mapping from the Euclidean space to the Minkowsky space is also 
illustrated.

\end{abstract}
\pacs{}
\maketitle

\section{Introduction}

\par
Here I solve directly in Minkowsky space the mass gap equation for 
Spontaneous Chiral Symmetry 
Breaking (S$\chi$SB). 
S$\chi$SB was introduced by the original work of Nambu and Jona-Lasinio
\cite{Nambu},
and it is presently accepted to occur in hadronic physics,
where it generates a constituent mass for the quarks.  S$\chi$SB also
implies the Partially Conserved Axial Current theorems
\cite{PCAC}, 
including a small mass for the pion. 
In the literature the mass gap equation is usually solved either 
in equal time
\cite{Instantaneous} 
or in Euclidean space
\cite{Euclidean},
in order to avoid the poles and complex quantities which are
expected in full Minkowsky calculations. 
Recently the scientific community has been exploring different
approaches to the Minkowsky space 
\cite{Minkowsky}.
An exact solution in Minkowsky space will test the quality of the 
aproximate solutions.
Moreover a solution in Minkowsky space opens wider applications.
For instance the Bethe Salpeter amplitudes can be computed on the mass shell
momentum $p^2=M^2>0$,
and at the same token it is possible to boost the hadrons to any convenient
frame.
In this paper an analytical approach is applied to finite and analytic kernel.
For simplicity the infrared or ultraviolet divergences are not addressed here.
Although a complete solution can only be achieved numerically, an interesting 
insight is nevertheless presented by the analytical approach. 

\par
In the Schwinger-Dyson formalism truncated at the rainbow level, 
the mass gap equation is,  
\FL
\begin{eqnarray}
{\cal S}^{-1}(p)&=&{\cal S}_{0}^{-1}(p)\ - \ \
\begin{picture}(60,35)(0,0)
\put(0,5){
\begin{picture}(60,25)(0,0)
\put(24,3){$_{S(q)}$}
\put(7.5,15){$_{-i V(p-q)}$}
\put(-9,2){$_{\gamma}$}
\put(55,2){$_{\gamma}$}
\put(55,-2){\vector(-1,0){36}}
\put(55,-2){\line(-1,0){60}}
\put(50.00,0.00){$\cdot$}
\put(49.69,3.92){$\cdot$}
\put(48.77,7.72){$\cdot$}
\put(47.27,11.84){$\cdot$}
\put(45.22,14.40){$\cdot$}
\put(42.67,17.68){$\cdot$}
\put(39.69,20.22){$\cdot$}
\put(36.34,22.28){$\cdot$}
\put(32.72,23.78){$\cdot$}
\put(28.91,24.70){$\cdot$}
\put(25.00,25.00){$\cdot$}
\put(21.08,24.70){$\cdot$}
\put(17.27,23.78){$\cdot$}
\put(13.65,22.28){$\cdot$}
\put(10.30,20.22){$\cdot$}
\put(7.32,17.68){$\cdot$}
\put(4.77,14.40){$\cdot$}
\put(2.72,11.84){$\cdot$}
\put(1.22,7.72){$\cdot$}
\put(0.30,3.92){$\cdot$}
\put(0.00,0.00){$\cdot$}
\put(48,-4){$\bullet$}
\put(-2,-4){$\bullet$}
\end{picture} }
\end{picture} \ ,
\nonumber \\
\vspace{1cm}
S(p)&=&{i \over A(p^2)\not{p}-B(p^2)+i\epsilon } \ .
\label{propagator}
\end{eqnarray}
For the quark-quark interaction I consider the class of kernels that can
be decomposed in a class of Yukawa potentials, which 
is finite both in the ultraviolet and infrared limits,
\begin{equation}
-i \,V(p-q)=-i \ \sum_i { \alpha_i \ 4 \pi \over (p-q)^2- \lambda_i^2+i\epsilon} \ .
\label{kernel}
\end{equation}
To ensure that the kernels are finite in the ultraviolet limit of large $(p-q)^2$ 
I assume that $\sum_i \alpha_i =0 $, since this  implies that
the kernel vanishes at least like $[(p-q)^2]^{-2}$. In the infrared limit the
kernels are finite when the $\lambda_i$ are finite, or when infrared cancellations
occur. 
The class of kernels defined in eq. (\ref{kernel}) is quite general, for instance it 
includes the Coulomb interaction regularized by a Pauli-Villars term. It also includes 
the Fourier transform of the confining linear potential, 
\begin{equation}
F. T. \left(|{\bf r}| e^{-\lambda |{\bf r}| } \right)
= {d^2 \over d\lambda^2 }  {4 \pi \over {\bf k}^2+\lambda^2 } \ ,
\label{IRfinite}
\end{equation}
and of the linear potential with a negative infinite shift
\begin{equation}
F. T. \left(-e^{-\lambda |{\bf r}| } \right)
= {d \over d\lambda }  {4 \pi \over {\bf k}^2+\lambda^2 } \ ,
\label{IRdivergent}
\end{equation}
which are both obtained in the limit of vanishing $\lambda$. Using finite 
differences, the derivatives in $\lambda$ can be decomposed in the class of 
potentials of eq. (\ref{kernel}). The class of kernels addressed in this paper 
is not only general, it is also covariant, analytical and forward propagating, 
and this is convenient to study the mass gap equation in Minkowsky space.

\par
Chiral symmetry breaking occurs when a non-vanishing mass is dynamically
generated. For clarity let us also assume that $A\simeq 1$ in eq. (\ref{propagator}).
This approximation is qualitatively acceptable when the kernel is finite
\cite{Instantaneous, Euclidean}.
Computing $A$ is not difficult, but it obscures the result of the paper.
Then the mass gap equation is a single non-linear and integral equation for $B=M$,
\begin{equation}
\hspace{-.1cm}M(p^2) = m_0+\int_{-\infty}^\infty  {i \, d^4 q \over (2 \pi)^4}
 V(p-q) {  M(q^2)\over q^2-M^2(q^2)+i\epsilon} \ ,
\label{mass equation}
\end{equation}
where all Dirac and colour algebraic factors are absorbed in the coupling constants
$\alpha_i$ of the potential $V(p-q)$.
The chiral limit of vanishing current quark mass (the mass in the free quark propagator)
$m_0\simeq 0$, is also assumed. This is particularly interesting because it applies to 
the physics of the quarks $u$ and $d$. 
It is clear that eq. (\ref{mass equation}) then has a trivial solution $M(p^2)=0$.
The problem that this paper addresses is the other possible solutions of 
eq. (\ref{mass equation}), with the kernel of eq. (\ref{kernel}).
The technical difficulties reside in the multiple integral with poles and 
complex quantities and in the non-linearity of the self-consistent equation.

This paper is organised in Sections. In Section \ref{standard} standard 
approximations are applied to the the rainbow Schwinger-Dyson, and the need to
perform a calculation in Minkowsky space is motivated. In Section \ref{one loop} 
the integral in the mass gap equation is computed analytically in the
case where the quark mass is assumed to be constant. The solution of the 
Minkowsky non-linear integral mass gap equation is addressed in Section
\ref{non-linear}. Finally In section \ref{conclusion} the results and conclusion
are presented.

\section{Using standard approximations}
\label{standard}

In the literature the mass gap equation is usually solved either in equal time
\cite{Instantaneous} 
or in Euclidean space
\cite{Euclidean}. 
In the equal time approximation, the Lorentz invariance is lost.
The space and time components of physical constants, say ${f_\pi}^s$ or
${f_\pi}^t$ may differ
\cite{fpi}.
Moreover it is not clear how to boost the hadrons outside the centre of mass frame.
In the Euclidean approximation it is not clear if a simple Wick rotation 
$p_0 \rightarrow i\,p_4$ is exact because there may exist poles 
in the path ot the $p_0$ axis. Moreover it is very hard to rotate
back to Minkowsky with the inverse rotation $i\,p_4 \rightarrow p_0$
when only a numerical expression of the dynamical mass is known. 
This is connected to the topological problem of discretising a curved 
surface.
Although these two methods are approximate, they are used in the literature because 
they are fully consistent with dynamical symmetry breaking, providing the same 
approximation is also used in the bound state equations
\cite{PCAC}. 
At the same token these methods avoid the technical problem of addressing poles 
and complex quantities, which are expected in Minkowsky space.
By providing a solution of the mass gap equation, both Lorentz invariant
and in Minkowsky space, both the equal time and the Euclidean time 
approximations may be better understood. Here the approximate methods for
solving the Schwinger-Dyson equation are reviewed.


\par
In the instantaneous or equal time approximation,
it is assumed that the dependence in 
$p_0$ and in $q_0$ is irrelevant in the kernel of the
mass gap equation. 
The angular integral of the coulomb
potential with neglected time component is,
\begin{eqnarray}
&&\int_{-1}^1 d\omega { 2 \pi {\bf q}^2 \, \alpha_i \over {\bf p}^2 
+{\bf q}^2 -2 |{\bf p}| |{\bf q}| \omega+ \lambda_i^2 }
\nonumber \\
&=& { 2 \pi {\bf q}^2 \, \alpha_i \over -2 |{\bf p}| |{\bf q}| } \log 
\left[ (|{\bf p}|-|{\bf q}|)^2 + \lambda_i^2\over  (|{\bf p}|+|{\bf q}|)^2 
+ \lambda_i^2\right] \ ,
\label{Instantaneous kernel}
\end{eqnarray}
where the notation for quadri-vectors and for tri-vectors is
$p=(p_0,{\bf p})$.
In what concerns the mass, a solution exists where the
mass is independent of $q_0$, and the integral in
$q_0$ is trivial,
\begin{eqnarray}
\int_{-\infty}^{\infty} dq_0
{ i \over {q_0}^2 -[ {\bf q}^2 + M^2({\bf q}^2) - i \epsilon ]}
= { \pi \over  \sqrt{ {\bf q}^2 +M^2( {\bf q}^2 ) } }
\end{eqnarray}
and the mass gap equation in the instantaneous approximation is,
\begin{eqnarray}
M( {\bf p}^2) &=& \int_0^\infty d |{\bf q }| \sum_i 
{ - \alpha_i \, |{\bf q}| \over 4 \pi |{\bf p}| } \log 
\left[ (|{\bf p}|-|{\bf q}|)^2 + \lambda_i^2\over  (|{\bf p}|+|{\bf q}|)^2 
+ \lambda_i^2\right] 
\nonumber \\
&&
{ M( {\bf q}^2) \over  \sqrt{ {\bf q}^2 +M^2( {\bf q}^2 ) } } \ .
\label{MGE instantaneous}
\end{eqnarray}
This is a one-dimensional non-linear integral equation. It has no singularities and 
it is solvable numerically
\cite{Instantaneous}.
 

\par
In the Euclidean approximation, it is assumed that the time component
of the momentum can be replaced $p_0\rightarrow i\,p_4$ or 
$-{p_0}^2\rightarrow {p_4}^2$ .
A convenient angular description of the variables is,
\begin{equation}
\left\{
\begin{array}{cccc}
p_1=p_E & \sin \phi & \sin \theta & \sin \eta \\
p_2=p_E  & \cos \phi & \sin \theta & \sin \eta \\
p_3=p_E  & & \cos \theta & \sin \eta \\
p_4=p_E  & & & \cos \eta 
\end{array}
\right.
\label{Euclidean angles}
\end{equation}
where the Euclidean momentum is,
$ {p_E}^2= -p^2={p_1}^2+{p_2}^2+{p_3}^2+{p_4}^2 $.
In this case the boson exchange potential of eq. (\ref{kernel})
is finite for all momenta. Covariance allows the choice of the
external momentum parallel to the fourth axis, and three angular
integrals can be performed,
\begin{eqnarray}
&& 
{ -4\pi \, \alpha \, {q_{\small E}}^3 \over 2 q_E }
\int_0^\pi d\eta { 
\sin^2 \eta \over {p_E}^2 +{q_E}^2 -2 {p_E} {q_E} \cos \eta + \lambda^2 }
\nonumber \\
&=& 2\pi^2 \alpha \, q^2 
{ p^2 +q^2 - \lambda^2 + \sqrt{ ( p^2 +q^2 - \lambda^2 )^2 - 4 p^2 q^2 }
\over 4 p^2 q^2 }
\nonumber \\
&=& { \pi^2 \alpha \over 4 {p}^2 }
\left[\sqrt{  -(p +q)^2 + \lambda^2 }-\sqrt{  -(p -q)^2 + \lambda^2 }\right]^2
\label{Euclidean kernel}
\end{eqnarray}
which is a function of the real and positive variable 
$- q^2={q_E}^2$.
Again the mass gap equation is reduced to a one-dimensional non-linear integral equation, 
\begin{eqnarray}
M(p^2)&=&\int_0^\infty d(-q^2) \sum_i 
{ \alpha_i  \over 2 \pi p^2 } \Bigl[ 
 p^2 +q^2 - \lambda^2 +
\nonumber \\
&& \hspace{-1cm} \sqrt{ ( p^2 +q^2 - \lambda^2 )^2 - 4 p^2 q^2 }
\Bigr] 
{ M(q^2) \over -q^2 + M^2(q^2) }
\label{MGE Euclidean}
\end{eqnarray}
which has no singularities and is solvable numerically 
\cite{Euclidean}.


\par
A third perspective, differing from the equal-time approximation because
it includes the retardation in the kernel and differing from the Euclidean
approximation because it addresses time-like momenta, is provided by the coupled 
channel approach. Here the approximation consists in only considering the positive energy
poles of the propagators. For simplicity I consider here a single Yukawa term,
\begin{eqnarray}
&&{1 \over q^2 -M^2 +i \epsilon } \rightarrow 
{1 \over q_0- H_f +i \epsilon } 
{1 \over 2 H_f} 
\nonumber \\
&&{1 \over (p-q)^2-\lambda_i^2 +i \epsilon } \rightarrow 
{1 \over E- q_0- H_b +i \epsilon } 
{1 \over 2 H_b } 
\nonumber \\
&& H_f({\bf q})= \sqrt{{\bf q}^2 +M^2} 
\nonumber \\
&& H_b({\bf p-q})= \sqrt{({\bf p-q})^2 +\lambda_i^2} 
\end{eqnarray}
then the integral in $q_0$ produces,
\begin{eqnarray}
\int{dq_0 \over 2 \pi}
{i \over q_0- H_f +i \epsilon } 
{i \over E- q_0- H_b +i \epsilon } 
\nonumber \\
= {i \over E-H_f - H_b+i \epsilon  }
\end{eqnarray}
The $1\over 2 H_f $,  $1\over 2 H_b $,  and the fermion mass $M$ can be 
included in the boson creation vertex,
\begin{equation}
\Gamma^\dagger = \sqrt{M \over 4 \,  H_f \, H_b} \ ,
\end{equation}
then the mass gap equation is  equivalent to a coupled channel Hamiltonian equation,
where the one fermion channel is coupled to the fermion plus
boson channel by the boson creation and annihilation vertices,
\begin{equation}
\left[
\begin{array}{cc}
H_f -i\epsilon -E & \Gamma  \\
\Gamma^\dagger & H_f +H_b -i\epsilon -E
\end{array}
\right]
\left(
\begin{array}{c}
\phi_f \\
\phi_{f,b}
\end{array}
\right)
=0 \ .
\end{equation}
Reducing the equation by substitution of the fermion plus boson 
wave-function, the secular equation is,
\begin{eqnarray}
0&=&\Bigl[E - H_f(p) - \int { d^3 p \over (2 \pi)^3} \Gamma (p,q) 
\nonumber \\
&& \hspace{-1cm}
{ 1 \over E- H_f(q) -H_b(p-q) + i \epsilon } 
\Gamma^\dagger(p,q) \Bigr] \phi(p) = 0 \ ,
\label{MGE secular}
\end{eqnarray}
where it is clear that when $ E > M + \lambda_i$, the threshold
for a boson production is open. This produces a cut in the function
$M({\bf p})$, with an imaginary component for the mass. Actually
more cuts appear when the energy is further increased, and the
system couples to to 1 fermion and 2 bosons,  1 fermion and 3
bosons ... Therefore expect cusps and imaginary components are
expected to appear when $ E= M +\lambda_i $, $E= M +2 \lambda_i$ ...

\par
When the mass gap equation is solved in the Minkowsky space, then all the features
of the three different approaches of eqs. 
(\ref{MGE instantaneous}), (\ref{MGE Euclidean}) and (\ref{MGE secular}) are 
expected to appear. Non-trivial solutions of the mass gap equation, and 
imaginary masses above thresholds are expected. Moreover It will be interesting to 
study the effect of the negative energy components on eigenvalues.

\section{One loop analytical calculation}
\label{one loop} 

\par
As a first approach to the Minkowsky integral, I assume a real constant mass
$M$ and compute analytically the integral of eq.(\ref{mass equation}). 
This can be regarded a one loop approximation to the mass gap equation, in the 
sense that the correct solution can be obtained iteratively, in an infinite
loop calculation. The analytical result may also be eventually used to remove 
poles form the numerical iterative program. In this sense I choose to perform
the multiple integrals in the same ordering that may be used in a
numerical integration. 
A first attempt to start by an analytical integral of the three angular integrals 
(both trigonometric and hyperbolic) was abandoned because the result was 
indeterminate. 
Therefore the two angular integrations of the three dimensional space are first 
performed, in order to simplify the kernel. Finally the double integration in the 
temporal $q_0$ and in the spatial $q$ are performed. 

\par
The main aim of this section is the covariant study of the 
positive $p^2$ case which was not accessible to the approximate approaches.
Covariance allows one to consider the $p=(p^0, {\bf 0})$ case. Analyticity 
can be used later, to continue this function to a $p^2$ negative case. The two 
dimensional angular integrals are trivial. The integral in $q_0$ is also 
directly computed, using the residue theorem,
\begin{eqnarray}
I(p_0,0) &=& 
\int d^3 {\bf q} d q_0
\sum_i { \alpha_i \over (q_0-p_0)^2-{\bf q}^2-{\lambda_i}^2 + i\, \epsilon} 
\nonumber \\   
&&
{ i {\bf q}^2 \over {q_0}^2-{\bf q}^2-M^2+i\epsilon}
\nonumber \\
&=&
\int d |{\bf q}|  
\sum_i {- \pi  4 \pi {\bf q}^2 \over 2 
\sqrt{{\bf q}^2 + \lambda_i^2-i\epsilon}\sqrt{{\bf q}^2 + M^2-i\epsilon}}
\nonumber \\
&&  \left( {1 \over p_0 +\sqrt{{\bf q}^2 + \lambda_i^2-i\epsilon} + 
\sqrt{{\bf q}^2 + M^2-i\epsilon}} \right.
\nonumber \\
&& \left. + {1 \over -p_0 +\sqrt{{\bf q}^2 + \lambda_i^2-i\epsilon} + 
\sqrt{{\bf q}^2 + M^2-i\epsilon}}
\right) \ .
\label{first integral}
\end{eqnarray}
The integral in $q=|\vec q |$ can also be performed analytically. It is interesting
to remark that the threshold for an imaginary contribution appears when
$p_0>M+\lambda_i$. Indeed the pole appears at the root $\rho$ of,
\begin{eqnarray}
p_0&=&\sqrt{{\bf q}^2+M^2}+\sqrt{{\bf q}^2+\lambda_i^2} \Rightarrow |{\bf q}|= \rho ,
\\ \nonumber
\rho &=& 
\sqrt{ \left(p_0^2 -M^2 - \lambda_i^2\right)^2 -4 M^2 \lambda_i^2 \over 4 p_0^2 } 
+ i \epsilon \ 
\\ \nonumber 
\label{pole}
\end{eqnarray}
where $\rho$ coincides with the mass shell momentum above threshold.
A simple algebraic simplification transforms the integral into,
\begin{equation}
I= \int d|{\bf q}| {- \pi \over 4 p_0^2} { 4 \pi {\bf q}^2 \over {\bf q}^2 - \rho^2} \left( 
  { M^2 + p_0^2 - \lambda_i^2 \over \sqrt{M^2 + {\bf q}^2}} + 
{-M^2 + p_0^2 + \lambda_i^2 \over \sqrt{{\bf q}^2 + \lambda_i^2}} \right) \ .
\label{pole separated}
\end{equation}
where the pole $\rho $ defined in eq. (\ref{pole})
is real when the threshold opens, at $p_0>M+\lambda_i$.
$\rho$ is also real when $p_0<|M-\lambda_i|$, below
the pseudo-thershold, 
\begin{equation}
\rho= 
{ \sqrt{ \left[p_0^2 -(M + \lambda_i)^2 \right] \left[p_0^2 -(M - \lambda_i)^2 \right] 
\over 4 p_0^2}}
+ i \epsilon \ \ ,
\end{equation}
however in this case the residue vanishes,
therefore $I$ is only expected to possess an imaginary component above the
threshold for the fermion-boson production.

\par
\begin{figure}[b]
\begin{picture}(250,170)(0,0)
\put(0,0){\epsfig{file=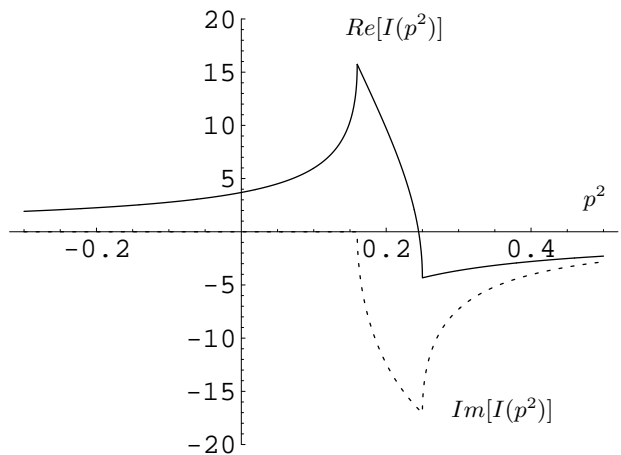,width=9.cm}}
\put(140,160){$ Re[I(p^2)] $}
\put(180,15){$ Im[I(p^2)] $}
\put(230,95){$ p^2 $}
\end{picture}
\caption{ Real and imaginary part of the integral $I( p^2 )$ that dynamically 
generates quark mass. Here the parameters $ M=0.3$ GeV , $ \lambda_1=0.1$ GeV  and 
 $ \lambda_2=0.2$ GeV, $\alpha_1=1$ and $\alpha_2=-1$  are used. 
The cusps occur precisely where the thresholds open, at $p^2= M+\lambda_1$ 
and at $p^2= M+\lambda_2$. }
\label{real and imaginary}
\end{figure}

After cumbersome calculations, where the ultraviolet divergent terms
cancel because they are proportional to $\sum_i \alpha_i$, the
integral $I(p_0,0)$ can finally be reduced to the exact form,
\begin{widetext}
\begin{eqnarray}
I(p^2)&=&\sum_i {\alpha_i \, \pi^2  \over 2 p^2 }
\Biggl[ \left(-M^2 + p^2 + \lambda_i^2\right)
\log {  M^2 \over \lambda_i^2} 
\nonumber \\
&& +\left(-M^2 + p^2 + \lambda_i^2\right)
\Biggl( 
{ \rho \over \sqrt{ \rho^2 + \lambda_i^2}} \Biggl\{ 
 \log\left[ i \, \left( 1 - { \sqrt{\rho^2 + \lambda_i^2} \over \rho} \right) \right] - 
       \log \left[ i \, \left(1 + { \sqrt{\rho^2 + \lambda_i^2}\over \rho} \right) \right] 
\Biggr\} \Biggr)
\nonumber \\
&&
+ (M^2 + p_0^2 - \lambda_i^2)
{ \rho \over \sqrt{\rho^2 + M^2} }
    \Biggl\{ \log \Biggl[ 
i \, \left( 1 - { \sqrt{\rho^2 + M^2} \over \rho} \right) \Biggr]  
- 
\log \left[ i \, \left( 1 + { \sqrt{\rho^2 + M^2} \over \rho} 
\right) \right]  \Biggr\} \Biggr) \Biggr] \ ,
\label{analytical}
\end{eqnarray}
\end{widetext}
where $\rho$ is defined in eq. (\ref{pole}).  The result of eq. (\ref{analytical}) is 
not only correct for time-like momentum, it also applies to negative $p_0^2$,
where the integrals of eq. (\ref{first integral}) remain correct. 
The integral $I(p^2)$ is depicted in Fig. \ref{real and imaginary}.

\par
There are three particular cases of eq. (\ref{first integral}) that can be 
tested independently. The imaginary part can be directly computed with the 
residue theorem applied to eq.(\ref{pole separated}). 
When $p_0>M+\lambda_i$ the imaginary part is,
\begin{equation}
Im[I(p_0,0)]=\sum_i -\alpha_i \, 
\pi^3{\sqrt{\left(p_0^2 -M^2 - \lambda_i^2\right)^2 -4 M^2 \lambda_i^2 }
\over 
   p_0^2} \ .
\label{imaginary}
\end{equation}

%

Another important particular case is the matching point, between the time-like
momentum and the space-like momentum, of $p^2=p_0^2=0$.
The integral of eq. (\ref{first integral}) is then easy to compute, 
\begin{eqnarray}
I(0,0) =
\sum_i 
- \ \alpha_i \pi^2{ M^2 \log(M^2) - \lambda_i^2  \log(\lambda_i^2)
\over M^2 - \lambda_i^2 } \ .
\end{eqnarray}

Finally in the space-like case $p=(0,\bf p)$ can be considered. This implies
that the poles are on the correct quadrants to enable a trivial Wick rotation.
I can also use the angular integrals already performed in eq. (\ref{Euclidean kernel}). 
The resulting integral can be computed analytically,
\begin{eqnarray}
&&I(0,{\bf p}) = \sum_i
{- \alpha_i \, \pi^2 \over 2{\bf p}^2 }
\Biggl\{ 
   (M^2 + {\bf p}^2-\lambda_i^2)  
\log\left({\lambda_i^2 \over M^2}\right) + 
\nonumber \\
&&
\sqrt{({\bf p}^2 + M^2 + \lambda_i^2)^2 - 4 M^2 \lambda_i^2} \log \Biggl[ 
\\ \nonumber 
&&
{M^2 + {\bf p}^2 + \lambda_i^2 + 
\sqrt{ (M^2 + {\bf p}^2 + \lambda_i^2)^2-4 M^2 \lambda_i^2 }
\over
M^2 + {\bf p}^2 + \lambda_i^2 - 
\sqrt{ (M^2 + {\bf p}^2 + \lambda_i^2)^2-4 M^2 \lambda_i^2 }
}\Biggr]\Biggr\} \ .
\label{space-like}
\end{eqnarray}
These three particular cases comply with eq. (\ref{analytical}).

\section{Approximate analytical solution}
\label{non-linear}

\par
\begin{figure}[t]
\begin{picture}(250,170)(0,0)
\put(0,0){\epsfig{file=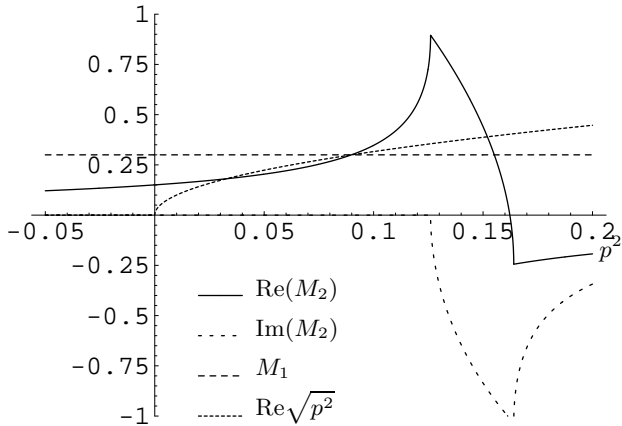,width=8.5cm}}
\put(230,65){$ p^2$}
\put(100,50){Re$(M_2)$}
\put(100,35){Im$(M_2)$}
\put(100,20){$M_1$}
\put(100,05){Re$\sqrt{p^2}$}
\end{picture}
\caption{ The initial mass $M_1$, the one loop mass $M_2(p^2)$ and
$\sqrt{p^2}$ are illustrated. They all coincide at 0.3 GeV,
and the parameters and $\alpha_1=-\alpha_2=36$ are adjusted accordingly. 
The parameters are $\lambda_1$=0.055 GeV and $\lambda_2$=0.105 GeV.
This case also includes a second solution of the mass gap equation,
$M^2=p^2=$ (0.18 GeV)$^2$. }
\label{double solution}
\end{figure}

Here the mass gap equation is solved in an analytical one loop 
approximation. The standard method to solve the mass gap equation 
(\ref{mass equation}) is the iterative method, where one starts by 
an initial educated guess for the mass $M_1(p^2)$. I consider
a constant $M_1$, which allows the analytical computation of the 
integral in eq. (\ref{mass equation}), with the techniques used in
section \ref{one loop}. This produces the next mass in the iterative
series,
\begin{equation}
M_1 \rightarrow M_2(p^2)= { 4 \pi \over ( 2 \pi )^4 } \, M_1 \, I(p^2) \ ,
\end{equation}
where the integral $I(p^2)$ is defined in eq. (\ref{analytical}).
To find an exact solution to the mass gap equation, one would then
need to continue this iterative process, computing again the integral
with the function $M_2(p^2)\rightarrow M_3(p^2)$ and so on until the 
method converges. 
However these further iterative steps would probably need a numerical 
computation, since the function $M_2(p^2)$ is already a complicated one. 
Therefore, in this analytical approach, I choose to stop at the second step 
of the iteration. To minimise the error of this one loop computation, I 
demand that the mass $M_2(p^2)$ coincides with the mass $M_1$ at the mass 
shell momentum  
$M_2(p^2)=p^2$, 
\begin{equation}
M_1 = M_2(p^2) = p^2 \ .
\label{demand}
\end{equation}
This is equivalent to assume that the mass dependence of the integral 
in eq. (\ref{mass equation}) is dominated by the pole neighbourhood.
The quark mass, dynamically generated in the mass gap equation, is 
expected to coincide with the constituent quark mass.
The constituent quark mass is estimated in the quark
model, where it is a crucial parameter to produce the hadronic spectrum,
and where it is of the order of 0.3 GeV.
Therefore the coupling constants $\alpha_i$ of the quark interaction 
(\ref{kernel}) are adjusted to reproduce $M_2=0.3$ GeV. An example
of this is illustrated in Fig. \ref{double solution}.
 
\par
I now discuss the different scenarios for the parameters 
$\lambda_i$ of the potential. I start with the case with {\em three 
Yukawa terms}. In the particular case of in eq. (\ref{IRfinite}),  
this corresponds to the linear potential, which vanishes in the infrared 
limit. It occurs that in this case the generated quark mass $M_2(p^2)$ 
is quite small close to the origin $p^2 \simeq 0$.
It is then very difficult to arrive at a quark mass of 0.3 GeV.
This may be related to the infrared cancellation,
\begin{equation}
\int { d^3 {\bf k} \over (2 \pi)^3}
 \left( {4 \pi \over {\bf k}^2+\lambda_1^2 } 
 -2 {4 \pi \over {\bf k}^2+\lambda_2^2 } 
 +{4 \pi \over {\bf k}^2+\lambda_3^2 } \right)
= 0 
\end{equation}
which occurs when the $\lambda_i$ are equally spaced,
$\lambda_3- \lambda_2= \lambda_2-\lambda_1$.
Moreover the structure of the potential is quite complicated 
above threshold, with large cusps. 
Therefore this class of models is abandoned.

%

\par
\begin{figure}[b]
\begin{picture}(250,170)(0,0)
\put(0,0){\epsfig{file=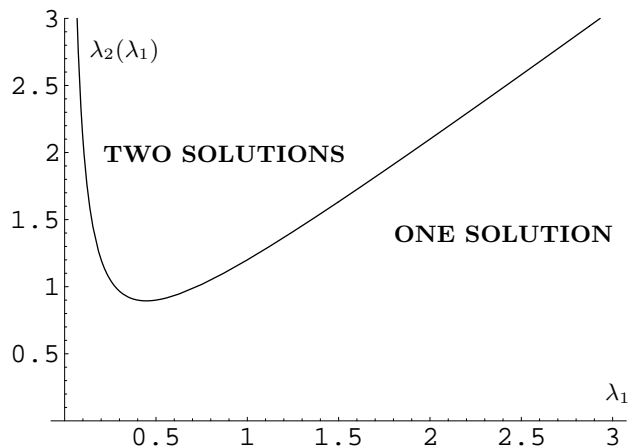,width=8.5cm}}
\put(35,150){$ \lambda_2(\lambda_1)$}
\put(230,20){$ \lambda_1 $}
\put(150,80){\bf ONE SOLUTION}
\put(40,110){\bf TWO SOLUTIONS}
\end{picture}
\caption{ Phases with one and two solutions of the mass gap equation
(\ref{mass equation}).}
\label{one and two solutions}
\end{figure}

Next the simpler case of {\em two Yukawa terms} is studied. This is related 
to the first derivative of eq. (\ref{IRdivergent}). In this case
it is easy to adjust the strength parameters $\alpha_1=-\alpha_2$ 
to produce a quark mass of 0.3 GeV at the pole position of $p^2=0.3$ 
GeV $^2$. 
This seems to agree with the infrared finite result,
\begin{equation}
\int { d^3 {\bf k} \over (2 \pi)^3} 
\left(
{4 \pi \over {\bf k}^2+\lambda_1^2 } 
 - {4 \pi \over {\bf k}^2+\lambda_2^2 } \right) 
= { \lambda_1-\lambda_2 \over 4 \pi} \ . 
\end{equation}
Nevertheless, in this simple case of two Yukawa terms,
there are two different scenarios. Depending on the
steepness of the function $M^2(p^2)$, it intercepts
the function $p^2$ either at a single point or in two
points. 
From a numerical exploration one concludes that
the two scenarios are separated by the line 
$\lambda_1(\lambda_2-\lambda_1) \simeq 0.2 $ GeV$^2$. This
line is depicted in Fig. \ref{one and two solutions},
in the parameter space $\lambda_2 \, \lambda_1$.
There is a single solution when 
\begin{equation}
\lambda_1(\lambda_2-\lambda_1) \geq 0.2 GeV^2 \ .
\end{equation}
For instance this includes the case of a large $\lambda_2$, which acts 
as an ultraviolet Pauli-Villars cutoff, say 
$ \lambda_1=0.3 GeV , \ \lambda_2=3 GeV $. 
When $ \lambda_1$ and $\lambda_2$ are quite 
large, the solution $M(p^2)$ of the mass gap equation is very 
smooth, and the Instantaneous or the Euclidean approaches constitute 
very good approximations to the actual Minkowsky solution. It is also 
clear that in the case of a Pauli-Villars regularisation $A\simeq 1$ 
is not acceptable. $A$ would need to be computed, nevertheless its 
computation could be performed with the techniques presented in 
this paper.

%

\par
One also concludes that there is a double solution when
\begin{equation}
\lambda_1(\lambda_2-\lambda_1) \leq 0.2 GeV^2 \ .
\end{equation}
This includes the case of small and similar $\lambda_i$. This case is related to the 
first derivative of eq. (\ref{IRdivergent}).In the limit of vanishing
parameters $\lambda_i$, eq. (\ref{IRdivergent}) corresponds to a linear potential 
with a negative and infinite constant shift,
\begin{equation}
-{\sigma \over \lambda } \, e^{-\lambda |{\bf r}| } 
\simeq { -\sigma \over \lambda} + \sigma {\bf r} \ .
\end{equation}
Comparing with the potential defined in eq. (\ref{kernel}) the string tension
is $\sigma= \alpha_1 (\lambda_2^2-\lambda_1^2)/2$. Assuming 
a string constant of the order of 0.14 GeV$^2$ estimated from
the quark spectrum, one arrives at the following parameters, 
\begin{eqnarray}
\alpha ( \lambda_2-\lambda_1)&=& 1.8 \, GeV \ , 
\nonumber \\
{ \lambda_1 + \lambda_2 \over 2} &=& 0.08 \, GeV \ ,
\end{eqnarray}
and this implies that the potential also includes a negative constant 
shift of the order of -1.7 GeV. An example of the generated masses is 
depicted in Fig. \ref{double solution}.

\par
To proceed iterating the equation (\ref{mass equation}), say at two loop order
or more, or to compute the quark condensate $\langle \bar \psi \psi \rangle$, 
the analytical integral becomes quite complicated. 
A numerical study is probably necessary and this is not addressed in this paper. 
Nevertheless the qualitative changes to the one loop computation can be anticipated. 
In what concerns the positive $p^2$, it is expected that new channels will open whenever 
$ p^2 > (M+n \lambda_1 +m\lambda_2)^2$, where $m$ and $n$ are positive integers. This will 
affect the imaginary part of the mass. In what concerns the behaviour of the
mass $M(p^2)$ for very large positive or negative $p^2$, an inspection
of eq. (\ref{mass equation}) shows that
\begin{eqnarray}
M(p^2)& \rightarrow & { 4 \pi \over g} 
{ \alpha_1 \lambda _2 -\alpha_2 \lambda_1 \over (p^2)^2 } \langle \bar \psi \psi \rangle
\nonumber \\
\langle \bar \psi \psi \rangle &=& g  \int_{-\infty}^\infty  {i \, d^4 q \over (2 \pi)^4}
 {  M(q^2)\over q^2-M^2(q^2)+i\epsilon} \ ,
\end{eqnarray}
and therefore the mass should vanish proportionally to $1/(p^2)^2$. In the
one loop approximation the mass only vanishes like $1/p^2$, see 
Fig \ref{real and imaginary}. Therefore the large momentum behaviour of the generated 
quark mass is expected to improve in the next iterations.

\section{Results and conclusion}
\label{conclusion}

I address the technically challenging problem of solving the
non-linear integral Schwinger-Dyson equation in full Minkowsky space.
An analytical approach is followed, and approximate but analytical
expressions for the quark mass are obtained. 

I find that the quark mass exhibits a branch cut above the
threshold for boson creation, including an imaginary 
component. In the case of a linear potential the mass gap equation 
is expected to have at least two solutions, and this agrees with
references \cite{replicas}.

The analytical continuation of a the numerical Euclidean space solution 
(with negative $p^2$) into the full Minkowsky has been studied in the 
literature
\cite{Minkowsky}.
Here I verify that this analytical continuation is not uniquely defined, 
even when a dense set of points is known in the the space-like $p^2<0$ sector. 
In particular the equation (\ref{mass equation}) includes at least three
vanishing imaginary numbers $-i \epsilon$, summed respectively to the 
masses $\lambda_1, \, \lambda_2$ and $M$. For external space-like momenta 
$p^2<0$, these $i \epsilon$ are irrelevant. However for time-like momenta
$p^2>0$ there are possible branch cuts both above threshold 
$p^2>(M+\lambda_i)^2$ and below the pseudo-threshold $0<p^2<(M-\lambda_i)^2$,
and the integral depends on the sign of each of the three vanishing $\epsilon$.
Only one integral, with all the three $\epsilon>0$ is causally correct. 
For instance in Fig. \ref{real and imaginary}
the correct imaginary part of the mass is negative, but a continuation 
with a positive imaginary would also be analytically possible. 
As another example, a naive extension of the Euclidean eq.
(\ref{space-like}) to $p^2>0$ fails to coincides with the full Minkowsky
solution eq. (\ref{analytical}) for some time-like momenta.

The next step of this study will consist in applying numerical methods
to proceed with the study of the mass gap equation. 
A numerical integration seems to be necessary to compute the quark condensate 
$\langle \bar \psi \psi \rangle$. 
Moreover the exact numerical solution of the mass gap equation 
(\ref{mass equation}) may also be attempted. 
With the analytical method of this paper, it may be possible to 
subtract the poles from the integrand and to compute their contribution
analytically.
Another interesting method may be the fully numerical Monte-Carlo 
method 
\cite{Monte Carlo}.
These numerical methods will be applied elsewhere.

\acknowledgements
I thank discussions with Frieder Kleefeld, Felipe Llanes-estrada
and with Kim Maung-Maung on the difficulties of continuing
analytically Euclidean space solutions to the Minkowsky space. 


%

\end{document}